\documentclass[acmsmall,monochrome]{acmart}
\usepackage{courier}
\usepackage{url}
\usepackage{graphicx}
\usepackage{color}
\usepackage{array}
\usepackage{subfigure}
\usepackage[T1]{fontenc}
\usepackage{aecompl}
\usepackage{multirow}
\usepackage{changepage}

\usepackage[all]{nowidow}
\newtheorem{resques}{Research Question}

\AtBeginDocument{%
  \providecommand\BibTeX{{%
    \normalfont B\kern-0.5em{\scshape i\kern-0.25em b}\kern-0.8em\TeX}}}

 \setcopyright{none}
\setcopyright{acmcopyright}
\copyrightyear{2020}
\acmYear{2020}
\acmDOI{00.0000/0000000}

\acmJournal{PACMHCI}
\acmVolume{0}
\acmNumber{0}
\acmArticle{0}
\acmMonth{0}

\begin{document}

\title{You Recommend, I Buy: How and Why People Engage in Instant Messaging Based Social Commerce}

\author{Hancheng Cao}
\authornote{The two first authors contributed equally and are ordered alphabetically.}
\email{hanchcao@stanford.edu}
\affiliation{
  \institution{Department of Computer Science, Stanford University}
  \city{California}
  \country{United States}}

\author{Zhilong Chen}
\authornotemark[1]
\email{czl20@mails.tsinghua.edu.cn}
\affiliation{
  \institution{Beijing National Research Center for Information Science and Technology (BNRist), Department of Electronic Engineering, Tsinghua University}
  \city{Beijing}
  \country{China}}

\author{Mengjie Cheng}
\affiliation{
  \institution{Harvard Business School}
  \city{Massachusetts}
  \country{United States}}

\author{Shuling Zhao}
\affiliation{
  \institution{Beijing National Research Center for Information Science and Technology (BNRist), Department of Electronic Engineering, Tsinghua University}
  \city{Beijing}
  \country{China}}

\author{Tao Wang}
\affiliation{
  \institution{Graduate School of Economics, Kyoto University}
  \city{Kyoto}
  \country{Japan}}
  
\author{Yong Li}
\email{liyong07@tsinghua.edu.cn}
\authornote{This is the corresponding author.}
\affiliation{
  \institution{Beijing National Research Center for Information Science and Technology (BNRist), Department of Electronic Engineering, Tsinghua University}
  \city{Beijing}
  \country{China}}

\renewcommand{\shortauthors}{Cao and Chen et al.}

\newcommand{\hancheng}[1]{{\textcolor{red}{ [hancheng: #1]}}}
\newcommand{\Magie}[1]{{\textcolor{blue}{ [Magie: #1]}}}
\newcommand{\zhilong}[1]{{\textcolor{orange}{ [Zhilong: #1]}}}
\begin{abstract}
As an emerging business phenomenon especially in China, instant messaging (IM) based social commerce is growing increasingly popular, attracting hundreds of millions of users and is becoming one important way where people make everyday purchases. Such platforms embed shopping experiences within IM apps, e.g., WeChat, WhatsApp, where real-world friends post and recommend products from the platforms in IM group chats and quite often form lasting recommending/buying relationships. How and why do users engage in IM based social commerce? Do such platforms create novel experiences that are distinct from prior commerce? And do these platforms bring changes to user social lives and relationships? To shed light on these questions, we launched a qualitative study where we carried out semi-structured interviews on 12 instant messaging based social commerce users in China. We showed that IM based social commerce: 1) enables more reachable, cost-reducing, and immersive user shopping experience, 2) shapes user decision-making process in shopping through pre-existing social relationship, mutual trust, shared identity, and community norm, and 3) creates novel social interactions, which can contribute to new tie formation while maintaining existing social relationships. We demonstrate that all these unique aspects link closely to the characteristics of IM platforms, as well as the coupling of user social and economic lives under such business model. Our study provides important research and design implications for social commerce, and decentralized, trusted socio-technical systems in general.
\end{abstract}



\maketitle

\section{Introduction}



{\color{blue}Instant messaging (IM) based social commerce is a notable business phenomenon emerging in recent years. This kind of social commerce heavily relies on IM, the type of online chat that offers real-time synchronous text transmission over the communication networks.} IM has become widely adopted for communication purposes since the 2000s \cite{shiu2004americans}, e.g., WeChat, WhatsApp, and IM based social commerce leverages existing social relationships and communication channels on IM for marketing. {\color{blue}IM based social commerce is growing rapidly, attracting hundreds of millions of users and penetrating people's shopping experiences and daily lives especially in China~\cite{Liu2015TAMBase}.} 
Take Pinduoduo, one of the largest IM based social commerce in China, as an example. Embedded in WeChat, the leading IM in China, Pinduoduo has acquired over 200 million users in less than three years. With its daily order volume ranking second in mainland China, it has posed considerable threats to traditional e-commerce giants like Alibaba, Amazon, and JD\footnote{https://www.forbes.com/sites/alexfang/2018/07/26/ipo-of-chinese-e-commerce-firm-pinduoduo-mints-new-young-billionaire/\#50da2a734024}. Similar scenarios are spotted on social commerce based on other IM platforms. WhatsApp based social commerce becomes increasingly popular, e.g., Meesho, which receives Facebook's first investment in India\footnote{https://techcrunch.com/2019/06/13/facebook-meesho-first-indian-startup-investment/}.

\begin{figure}[h]
  \centering
      \begin{subfigure}[An IM based social commerce sales group for parents where a mom is recommending items with words and pictures, illustrating how she had bought the item and how cheap it is that day]{
        \label{fig:interface1}
        \includegraphics[width=0.30\textwidth]{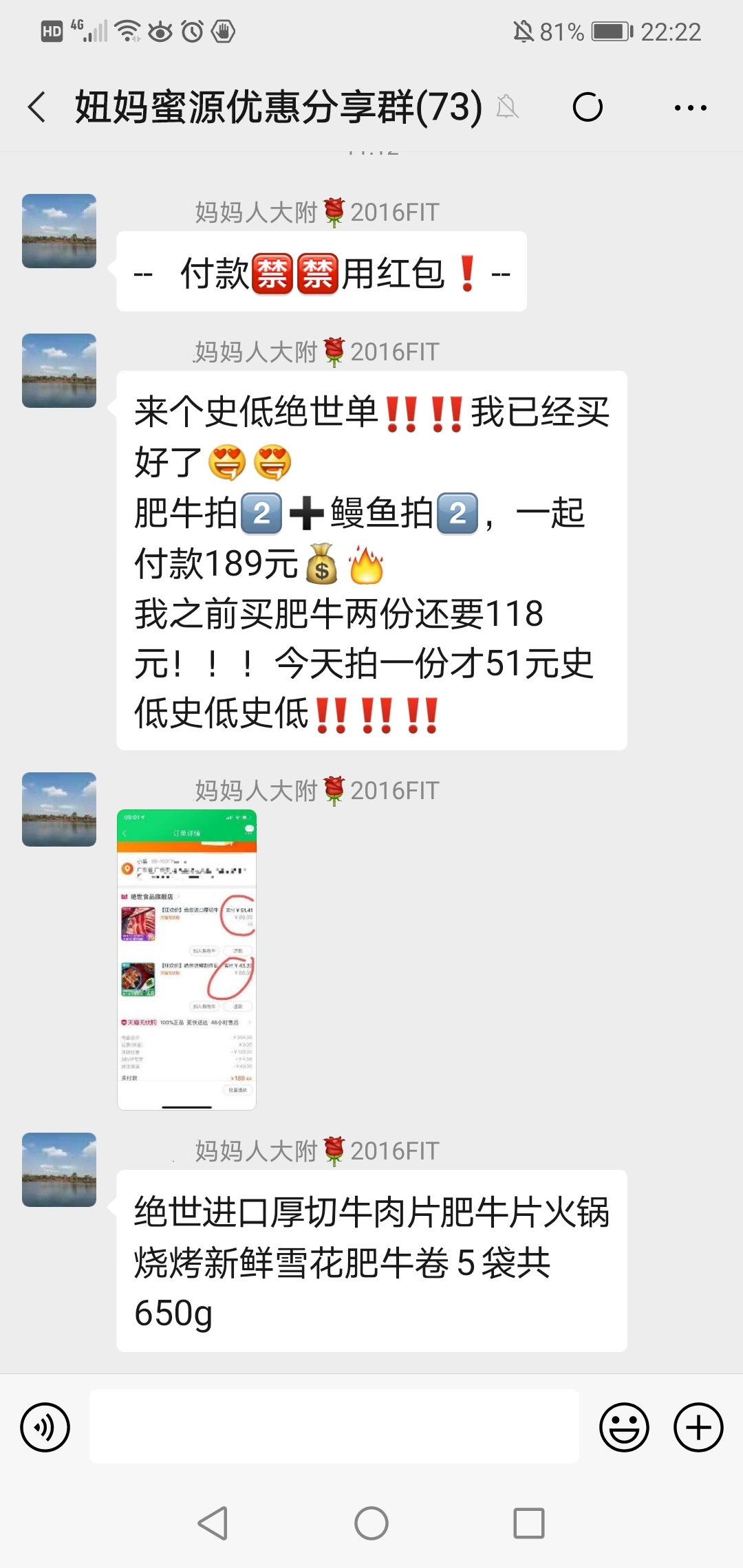}}
       \end{subfigure}
       \begin{subfigure} [An IM based social commerce sales group where group buyer are reporting their satisfying purchase experiences from the sales group with pictures and narrations]{
        \label{fig:interface2}
        \includegraphics[width=0.29\textwidth]{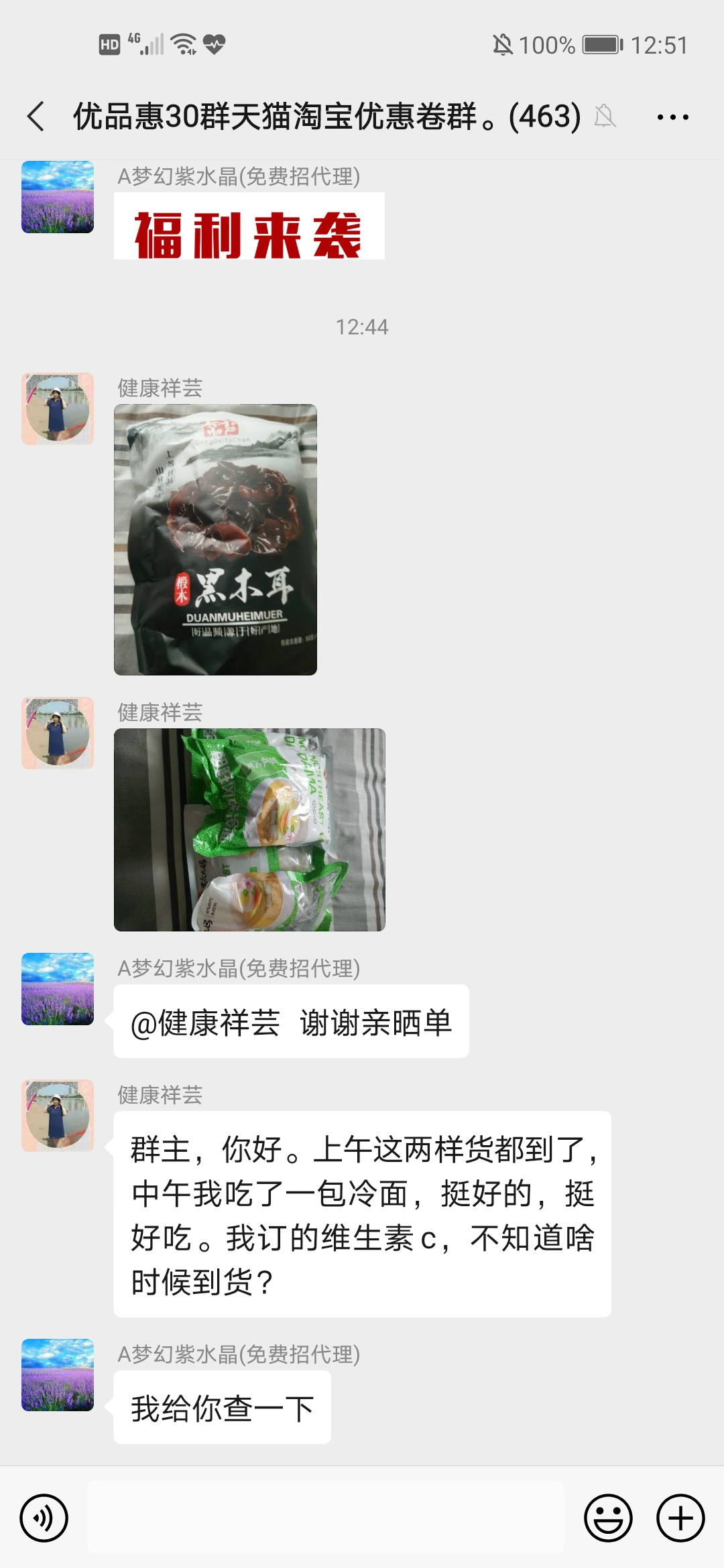}}
       \end{subfigure}

  \caption{Typical scenarios of IM based social commerce.}
  \label{fig:scenario}
    \vspace*{-4mm}
\end{figure}

As a recent instance of social commerce~\cite{liang2011introduction,zhou2013social,curty2011social} -- the act of using social media to promote online buying and selling of products and services
, IM based social commerce follows the tradition of incorporating economic transactions into social networks. Yet, in contrast to their prior counterparts which heavily rely on key opinion leaders such as celebrities and brands (e.g., Facebook commerce, Twitter commerce)~\cite{yadav2013social}, recent social commerce adopts a `decentralized' model~\cite{cao2019your} and leverages existing social relationships of ordinary people. Fig.~\ref{fig:scenario} demonstrates the typical scenario of IM based social commerce. It embeds shopping experience within IM, on which IM friends usually represent the users' real-world social network, \emph{i.e.}, people they know in real life~\cite{church2013s,wang_li_tang_2015}. {\color{blue}Specifically, on IM, group chats and point-to-point chats are the two most fundamental features. IM friends who share similar identity or have relatively close relationships typically socialize with each other through these group chats and point-to-point chats, forming groups ranging from family groups, classmates groups, co-worker groups, etc. IM based social commerce takes advantage of existing infrastructure on IM, where friends recommend and post products from the social commerce platforms in IM group chats (both groups specifically created for purchasing and ordinary group chats) or sometimes point-to-point chats, and quite often form lasting recommending/buying relationships.} As an important socio-technical system studied in the HCI and CSCW community, IM is known to demonstrate property deviating from other social platforms (e.g., Facebook, Twitter) in that it is more immersive, intimate, and strong-tie based~\cite{nardi2000interaction, hu2004friendships}. Therefore, it is likely that IM based commerce would give rise to unique user experiences unknown in the literature, which is essential to understand so as to better guide the platform's design and management. Yet so far few works have addressed the user behavior on emerging IM based social commerce beyond empirical measurement~\cite{cao2019your}. On the other hand, IM based social commerce can be seen as a recent large-scale `flexible' implementation of word-of-mouth marketing over strong ties. While traditional word-of-mouth marketing is mostly in person, over the telephone, or through email~\cite{silverman2011secrets}, IM based social commerce extend word-of-mouth over space and time, \emph{i.e.}, it is not restricted by physical distance and can act in both synchronous and asynchronous manners. It would thus be of interest to study how IM as a technology {\color{blue}bring new changes to} the word-of-mouth marketing practice.

{\color{blue}To fill in this research gap, in this work, we seek to understand user motivation and experience on IM based social commerce.} We focus our analysis on Chinese WeChat based social commerce, and aim to uncover how the implementation of social recommendation (act) on IM (socio-technical system) would result in unconventional experiences. Specifically, we ask: how and why do users engage in IM based social commerce? Do such platforms create novel shopping experiences that are distinct from prior commerce? And do these platforms bring changes to user social lives and relationships? 

To shed light on these questions, we launched a qualitative study where we carried out semi-structured interviews on 12 IM (WeChat) based social commerce users in China. Snowball sampling was adopted for interviewee recruitment and the interviews were conducted either face to face or remotely. Through analyzing the content mentioned by interviewees, we show that IM based social commerce does demonstrate significant characteristic differences compared to its prior counterparts. We discover that all unique aspects of IM based social commerce are closely linked to the embeddedness of user social and economic lives on the platform. Specifically, we demonstrate that IM based social commerce 1) enables more reachable, cost-reducing, and ubiquitous user shopping experiences, where users can buy cheaper, `decentralized' long-tailed products and experience novel products which otherwise they will not try or have never even heard of, largely as an effect of unique information diffusion mechanisms introduced by IM based social commerce through word-of-mouth, 2) {\color{blue}changes} user decision-making process in shopping due to pre-existing social relationships, mutual trust, shared identity, and community norm, which effectively addresses the issue of information overload in traditional commerce platforms, and 3) creates novel user social interactions, which contributes to new tie formation while maintaining existing social relationships. 

{\color{blue}Insights from our qualitative study on IM based social commerce can provide important research and design implications.} We reveal and discuss how ordinary people's social networks can achieve considerable economic value, and perhaps more importantly, shape the way idea flows and people interact, which gives birth to novel ways of interaction experience, as demonstrated by the case of IM based social commerce. We discuss how the unique ecosystem of IM helps marketing/economic transactions {\color{blue}to be} more engaging and inclusive, and how it enables more natural embeddedness of people's social and economic lives -- which are all in sharp contrast with prior key opinion leader based social commerce platforms. 
Finally, we discuss the potential downsides of such applications and how our research can help guide better design and management of social commerce and decentralized, trusted socio-technical systems in general.
\section{Background and Related Work}
In this section, we position our work in the literature from the following aspects: instant messaging platform (the technology platform our study focuses on), social recommendations (the activity we study), and the role of technology in this activity (agency).

\subsection{Instant Messaging Platforms}

Instant messaging (IM) is a type of online chat that offers real-time synchronous text transmission over the communication networks, which has become widely adopted especially for informal communication purposes since the 2000s \cite{shiu2004americans}. Nardi et al.~\cite{nardi2000interaction} show that IM not only supports flexible, expressive information exchange, but also enables the maintenance of a sense of connection with others and an active communication zone. Hu et al.~\cite{hu2004friendships} demonstrate that IM uses positively associate with verbal intimacy, affective intimacy, and social intimacy. Grinter and Palen indicate that IM has been a salient part of users' everyday experience and reflects real-life social relationships~\cite{grinter2002instant}. It has also been shown that on IM platforms, people tend to communicate more with each other when they share similar age, language, and location~\cite{leskovec2008planetary}, thus demonstrating homophily effect~\cite{mcpherson2001birds}. IM platforms have also been adopted in other scenarios, e.g., the workplace for collaboration and coordination~\cite{isaacs2002character, cao2021my}. What's more, IM as an interaction modality, has also been incorporated into various collaboration tools, e.g., remote meetings~\cite{cao2021large} and live streaming~\cite{chen2020large}.

The specific IM platform our paper focuses on (\emph{i.e.}, most Chinese IM based social commerce build upon) is WeChat, the most popular IM platform in China. It has several characteristics worth highlighting: 
a) \textit{Prevalence}. According to the latest report from Statista\footnote{https://www.statista.com/statistics/255778/number-of-active-wechat-messenger-accounts/}, there are over 1.2 billion monthly active users from a wide range of age groups in WeChat. Basically, one out of ten Chinese people use WeChat. 
b) \textit{Immersive}. WeChat is an IM platform embedded in daily life activities, including socializing, event planning, collaboration, and economic transactions (via WeChat Pay). It has also been used for more formal and professional communications, e.g., patient-provider communication~\cite{ding2019boundary}, organization-public engagement~\cite{tsai2018social}, etc. 
c) \textit{Strong ties}. Friends of WeChat are usually one's close social circle, either from work or in daily life, so the friend ties in WeChat are much stronger than many other communication platforms without real-life implications~\cite{wang2015dwelling, Cornell2016}. WeChat has some distinct features, e.g., the upper limit of group size in WeChat is 500 people and the posts on WeChat are only friend visible, which ensure private and strong-tie based communication experiences. In this work, we focus on the act of social recommendation/social commerce on WeChat, and study its implications on user experience and relationship and how it differs from prior counterparts.




\subsection{Social Recommendation and Word-of-Mouth Marketing}


Social recommendation~\cite{king2010introduction} - the utilization of social relations to recommend products or promote information, has been identified as a means to better unleash the power of social issues to benefit recommendation outcomes. For example, social influence~\cite{marsden1993network, sharma2016distinguishing} and homophily~\cite{mcpherson2001birds} have proven to be beneficial to recommendation performances~\cite{tang2013social,golbeck2006generating}. In practice, practitioners quite often leverage word of mouth for social recommendations, \emph{i.e.}, consumers share opinions, news, and information with their peers, which traditionally happens in person or over the telephone, but is made increasingly ubiquitous through email~\cite{bernstein2010enhancing}, social media platforms (e.g., Twitter, Facebook)~\cite{sharma2013friends}, and IM. 
Most of word-of-mouth marketing research focuses on the motivations~\cite{chung_darke_2006, word_of_mouth_1966, craig-lees_joy_browne_1995,lampel_bhalla_2007, sharma2015studying}, communication audience~\cite{reiss_2012, dubois_bonezzi_angelis_2016,argo_dahl_morales_2006}, communication channels~\cite{berger_iyengar_2013,duthler_2006}, etc., and it has been shown that social recommendation both helps people decide what to choose and provides social context that improves engagement~\cite{kulkarni2013all}. However, less attention is given to the effect of word-of-mouth marketing between offline and online (\emph{i.e.}, how word of mouth in real life among acquaintances and friends transfer into online purchase behavior) and its impact on the decision-making process. What's more, it is worth noting that even fewer works concentrate on the opposite direction of what changes can be brought by these marketing actions to real-life relationships~\cite{sharma2013popcore,caraway2017friends}. In this paper, we endeavor to uncover how the power of word of mouth can be leveraged on this new scenario of IM based social commerce, and how changes in the opposite direction, \emph{i.e.}, how actions of promotions shape real-life relationships, can be led to.

\subsection{Social Commerce}
Social commerce has gained increasing attention from both academics and practitioners recently, which implements social recommendation at scale over the Internet. Social commerce is defined as the use of Internet-based media to enable users to participate in the selling, buying, comparing, and sharing of information about products and services in online marketplace and communities~\cite{liang2011drives,zhou2013social}. 
Some scholars regard social commerce as a special kind of e-commerce, which allows the interaction between merchants and consumers in a social environment~\cite{sturiale_scuderi_2013, wu_shen_chang_2015}.

Social commerce varies in forms~\cite{curty2011social} and can be categorized into social network driven platforms (e.g., F-commerce, T-commerce), group buying platforms (e.g., Groupon), peer-to-peer sales platforms (e.g., eBay), and peer recommendation platforms (e.g., Amazon). Recently, there is also an increasing trend in utilizing social media platforms such as Instagram and Pinterest to do social commerce. For example, on Instagram, shoppable posts appear with little shopping bag icons in the left-hand corner of the image that users can click on to see the price and details about the product they are looking at. Pinterest creates the “Buyable Pins” feature to let users purchase products directly from pins and has begun positioning itself as less of a social media company but more of a commerce company since 2019\footnote{https://tamebay.com/2019/09/pinterest-social-commerce.html}. Compared to IM, the shopping behaviors happening on Pinterest and Instagram are more fun and exploratory, though these platforms seldom utilize the strong real-life social relationships that IM based platforms such as WeChat possess. Instead, they depend on key opinion leaders (e.g., celebrity, brand) or interest groups (e.g., people who share similar interests but quite often do not know each other in real life) to do the marketing.

Some earlier works in sociology and economics literature that studied the embeddedness of economic activities within social structure~\cite{granovetter1985economic}, and the interplay between market organization and trading relationships~\cite{geertz_1978} helps lay the theoretical foundation of social commerce research. Nowadays, increasing attention has been drawn to social commerce from many different disciplines ranging from computer science, economics, information system, and business over the years. Most of the works focus on the business model~\cite{cao2019your}, user behavior~\cite{kim2013effects,shin2013user,cai2014seller} and social network analysis~\cite{stephen2010deriving,holtz2017social}, etc. Complimenting these works, we seek to explore how people's experiences are shaped when the act of social commerce is combined with IM, where the social relationships delineate real-world social ties and thus the scenario better reflects the outcomes of social and economic blending.

\subsection{Our Focus: Emerging Instant Messaging (IM) based Social Commerce}


Based upon IM platforms (e.g., WeChat), recent years see the birth of a new type of social commerce on IM, which we refer to as IM based social commerce. Here shopping assistants on IM share product links in the IM based commerce group chats or one-on-one chats, motivated both by incentives from the platform (e.g., get discount when shopping on the platform) or by internal motivations (e.g., recommend friends products they find cost-efficient). Since IM based social commerce users offer useful product-related information through word-of-mouth and personal relationships, brands build trust among consumers. After the trust continues to deepen, consumer shopping behavior will extend from individuals to families and friends. Compared with traditional commerce, such IM based social commerce has the following advantages: 

a) \textit{Obtaining audiences more cost-effectively.} The network structure of IM based social commerce is rather decentralized, where ordinary people take the role of persuaders~\cite{cao2019your}. Leveraging the power of word-of-mouth on IM, IM based social commerce motivates the public to diffuse the promotions and stimulate viral cascades. As such, the monetary expenditure on engaging customers can be reduced: based on data from Analysys\footnote{https://www.sohu.com/a/331611239\_585973}, the customer acquisition cost of TMall increased by 60\% from 2015 to 2017 while JD increased by 164\% during the same period, both of which exceeded 250 RMB per person; but IM based social commerce platforms such as Perfect Diary can attract consumers to enter its WeChat pool with only 2-3 RMB for each.

b) \textit{Getting closer to customers.} Spreading under the context of WeChat, IM based social commerce leverages intimate friends and families around for promotions, entering into people's real-world social cycles~\cite{cao2019your,chen2020intermediary}. With close ties as mediums, an atmosphere of familiarity and comfort for interactions is created and the distance between brands and customers is reduced. What's more, the items promoted are also more friendly and approachable, where daily necessities take up a large share~\cite{chen2020intermediary}. 

c) \textit{Improving conversion rate.} Through the power of word-of-mouth, IM based social commerce enjoys the welfare of social influences. Through better matching, social enrichment, social proof, and price sensitivity mechanisms, it manages to attain significantly higher purchase conversion rates~\cite{xu2019think}.

We thus expect experiences on IM based social commerce would significantly deviate from other instances of social commerce. However, little work in the HCI and CSCW community studies user behavior and decision-making process on IM based platforms. The closest line of works focused on the usability of UI design~\cite{Minocha2006Evaluate, Rie2003Shiny,Stein2002Pic, H2002Interactive, Cai2008The}, choice overload on e-commerce platform~\cite{moser2017no}, features used by platforms to stimulate impulse buying~\cite{Moser2019Impulse}, etc. This work intends to bridge the gap and provide a better understanding of IM based social commerce. Here, we seek to investigate IM based social commerce from the standpoint of diffusion of innovation theory~\cite{rogers2010diffusion}: in our case, innovation is the product that is recommended, adopters are users who engage in buying, and the communication channel is the IM platform. This work thus focuses on how IM, as a novel communication channel for economic transactions, bring changes to user experience within this innovation diffusion process. Specifically, we propose to study the following research questions:

\begin{resques}[RQ\ref{resques:1}] \label{resques:1}
 How do people engage in IM based social commerce?
\end{resques}

\begin{resques}[RQ\ref{resques:2}] \label{resques:2}
How do shopping experiences on IM based social commerce differ from prior social commerce?
\end{resques}

\begin{resques}[RQ\ref{resques:3}] \label{resques:3}
 How does the embeddedness of social relationships and economic transaction influence user decision making on IM based social commerce?
\end{resques}

\begin{resques}[RQ\ref{resques:4}] \label{resques:4}
Does IM based social commerce bring changes to user social lives and relationships?
\end{resques}

\section{Method}
To answer our proposed research questions, we launched an interview study on IM based social commerce participators and buyers. We interviewed them either in person or through remote video/audio calls between January and April 2020. The interviews took the form of a semi-structured manner and probed into questions concerning how participants got involved in IM based social commerce, their experience on IM based social commerce, the reasons for and against IM based social commerce, changes introduced by IM based social commerce to their lives, etc. 

Following the grounded approach~\cite{corbin2014basics}, we interviewed and analyzed the content iteratively, consistently revising our interview protocol through the procedure so as to induce better themes. Before our interview recruitment, two of the authors joined, participated, and observed three WeChat groups where IM commerce is integrated for six months. We then consulted the administrators of these groups for participant recruitment recommendations and contacted the recommended participants with the group administrators' help. We utilized the recommended interviewees as seed participants and adopted the snowball sampling strategy~\cite{Biernacki1981Snowball} for further recruitment. The reason why we adopted the snowball sampling strategy is 1) to avoid the bias of the specificity and constraints of the groups we joined because the recommended person may be in different groups for IM commerce with distinct features, so that representativeness can be enhanced~\cite{sadler2010recruitment} through varying situations~\cite{strauss1998basics}; 2) to enable us to exert a certain degree of control on participant selection so that we can maximize differences and thus discoveries~\cite{strauss1998basics}; and 3) to ensure the quality of the interviews because recommended participants with inherent trust engendered among participants~\cite{sadler2010recruitment} are more likely to report their true experiences than random sampling. We did not cease the process of data collection until we reach theoretical saturation~\cite{corbin2014basics}, where no new themes emerge from the incorporation of new data. Eventually, 12 interviews were executed, among which 8 were taken face to face and 4 remotely. Table~\ref{tab:Interviewee} demonstrates the detailed information of the IM based social commerce buyers we interviewed. All the interviews were conducted in Mandarin and audio-taped after receiving participants' oral consent. The recordings of the interviews were further transcribed through the combination of transcription service and manual transcription. Identifiable information was removed to better protect participants' privacy. 


\begin{table}[!htbp]
\centering
\caption{Summary of the basic information of interviewees.}
\begin{tabular}{cccc|cccc}
\toprule
\textbf{ID} & \textbf{Gender} & \textbf{Age} & \textbf{\multirow{2}{*}{\shortstack{IM Purchase\\Frequency}}} & \textbf{ID} & \textbf{Gender} & \textbf{Age} & \textbf{\multirow{2}{*}{\shortstack{IM Purchase\\Frequency}}}\\
&&&&&&\\
\hline
P1 & F & 51-60 & High & P7 & F & 41-50 & Medium \\
P2 & F & 51-60 & High & P8 & M & 31-40 & Medium \\
P3 & F & 51-60 & High & P9 & F & 31-40 & Low \\
P4 & F & 51-60 & High & P10 & F & 31-40 & High \\
P5 & F & 41-50 & Medium & P11 & M & 31-40 & Low \\
P6 & M & 21-30 & Low & P12 & F & 21-30 & Medium \\
\bottomrule
\end{tabular}
\setlength{\belowcaptionskip}{10pt}%

\label{tab:Interviewee}
\vspace*{-4mm}
\end{table}

To analyze the interviews, we adopt the method of open coding~\cite{corbin2014basics}. Two Mandarin-speaking authors separately coded the first 10\% of the interview transcriptions and met and discussed the codes until agreements were made. Then, one of the first authors who is Mandarin-speaking coded the remaining transcriptions and termly discussed with three other Mandarin-speaking authors so as to guarantee consensus on the codes. Then, the codes and corresponding quotes are translated into English and two other Mandarin-speaking authors were responsible for validating the translations. When these steps were finished, the whole research team met and discussed thoroughly about the extracted content. Through sub-categorization and constant comparison~\cite{strauss1997grounded}, we developed and consistently revised the emerging themes and the following themes are established.

\section{Findings}
\subsection{How People Engage in IM Based Social Commerce}
We discover that people engage in IM based social commerce primarily through three channels: 1) standalone sales groups introduced by social ties, 2) sales groups emerged from existing social groups, and 3) existing social groups.

As a form of platforms that combine social relationships and economic interactions, IM based social commerce highly relies on people's social ties. With IM delineating people's intimacy where users' interactions are limited to those they know well or plan to know well~\cite{nardi2000interaction}, we find that the most common way people join IM based social commerce is through the introduction and recommendation by socially-close friends:

\begin{adjustwidth}{2em}{2em}
\emph{"I know her well. She is a parent of my son's schoolmates ... As soon as she hears that you have a need or you want to buy something, then she introduces you to the corresponding (WeChat) groups, where you head for different groups to buy different things."} (P3)
\end{adjustwidth}

Similar motivations are mentioned by P4, P6, P7, P8, P10, and P11, where introducers/referrers can range from classmates (P6), colleagues (P10), countrymen (P11), etc. These people's own satisfactory past shopping experiences  
are recognized as referrals and these experiences together with the enthusiastic recommendations drive users to a try. Through the power of word-of-mouth~\cite{westbrook1987product}, people get into these WeChat groups where in most cases the groups serve only for sales.

Another way people get involved in IM based social commerce is through sales groups that emerge from existing groups:

\begin{adjustwidth}{2em}{2em}
\emph{"It started out from WeChat groups, but not shopping groups at first. Perhaps they are, for example, ... parent groups whose children are students at [school name] ... And then many subgroups are divided from the groups, such as the flower and plant group, ... among which is a shopping group ... We join these groups based on our mutual interest."} (P2)
\end{adjustwidth}

Usually, the original groups are big enough to enable the formation of branches of sub-groups (P1, P2). However, not all groups can induce divisions. In most cases, groups where valid shopping subgroups can form are those whose members have the innate motivation to participate or are socially close enough, for example, \emph{"sharing the same identity"} (P2), and therefore are willing to engage in the sub-groups.

Sometimes these activities may occur directly in existing groups. For example, P1 mentioned she got involved in IM based social commerce through a mom group: 
\begin{adjustwidth}{2em}{2em}
\emph{"The mom group consists of the parents of my son's classmates in high school. After our kids graduate, some parents who get along well build a small group of approximately 20 people ... We talk about everything in the group ... (among which) if we find something good, ... we would share a link."} (P1)
\end{adjustwidth}

P2, P4, and P6 also note similar scenarios, but the group members may be colleagues (P4) and neighbors (P4 \& P6). In these cases, however, IM based social commerce is regarded as \emph{"recommendations rather than sales"} (P1). 

In summary, our findings further confirm that user engagement on IM based social commerce is heavily driven by existing social relationships on IM.

\subsection{User Experience on IM based Social Commerce}
In answer to RQ~\ref{resques:2}, we find that the foundational difference that distinguishes IM based social commerce from traditional e-commerce is that it provides much more inclusive and ubiquitous shopping experiences, both in terms of available product and engaging time. Specifically, with the benefit brought by IM, from the product angle, users are exposed to more diverse items that can fulfill their needs; from the perspective of time, users can more conveniently reach items at a broader time range. 

\subsubsection{Product-wise}
IM based social commerce enables participants to reach a wide range of products that are worthy of purchases. Specifically, IM provides a unique channel for certain kinds of product acquisition, making IM based social commerce stand out in the way that it helps users: 1) reach products in a cheap and cost-efficient way; 2) take more diverse and decentralized products into considerations; and 3) "jump out of the box" and try novel products.

\begin{adjustwidth}{0em}{0em}
\textbf{Cheap and cost-efficient.}
\end{adjustwidth}

Cheapness and cost-efficiency are frequently mentioned by participants as reasons for their purchases through IM based social commerce. For example, P7 and P1 mentioned they enjoyed purchasing seafood and underpants from IM based social commerce:

\begin{adjustwidth}{2em}{2em}
\emph{"One has to consider the quality, freshness, and price of the seafood ... it is generally rather expensive... but it is cheaper to buy through this method (IM based social commerce) ... I used to buy seafood at [supermarket name] where the quality can be ensured, ... but it is relatively expensive. The price is relatively low here (compared to [supermarket name])."} (P7)
\end{adjustwidth}
\begin{adjustwidth}{2em}{2em}
\emph{"You see the underpants I bought ... they are really cheap. They are only about 4 yuan for each pair, ... but if you buy them outside, normally they are more than 10 yuan."} (P1)
\end{adjustwidth}

Similar scenarios are also reflected by P2, P3, P5, P6, P9, and P10. In IM based social commerce, fewer layers lay between item and producer (P3). What's more, the so-called stores for marketing exist entirely in a virtual and online form. Therefore, the profits extracted by dealers at multiple levels can be substantially reduced and rent for physical stores can be saved. As such, users can more directly reach the products, and as a result, purchase the same products in a cheaper and more cost-efficient way.

\begin{adjustwidth}{0em}{0em}
\textbf{Diverse and decentralized.}
\end{adjustwidth}

Furthermore, IM based social commerce exposes people to longer-tailed products which are not well-known and popular among buyers, which leads to users' more diverse and decentralized shopping experience. P2 recalled her experience of purchasing daily nuts on IM based social commerce:

\begin{adjustwidth}{2em}{2em}
\emph{"I used to buy [brand name]'s daily nuts. The big brand. It is relatively expensive. Normally it costs about 2.5 yuan for each packet (on social commerce) ... but this costs less than 1 yuan for each packet ... I had never heard about the brand, but the contents are all nuts, real nuts, ... you know about the contents. It is very cost-efficient."} (P2)
\end{adjustwidth}


In the traditional setting of e-commerce, people often head directly to the targeting items~\cite{zeng2019user}, restricting the brand, price, etc. IM based social commerce is different in the way that the vast majority of the products that have been pushed and recommended are not those with "big" names. Although some of these products also exist on traditional e-commerce platforms, they fall short in the way that they are in lack of quality assurance, and are less likely to be recommended by recommendation algorithms. IM based social commerce deals with this through the innate social closeness and trust within people's social relationships. As mentioned by P2, 

\begin{adjustwidth}{2em}{2em}
\emph{"If my friends have tried the item, I would place an order without hesitation. ... If one has always been able to recommend something good, he will have some prestige and I will definitely trust his recommendation."} (P2)
\end{adjustwidth}

Building upon the reputation system embedded in social relationships, IM based social commerce makes items whose brand is not that famous but whose actual cost matches people's perceived value stand out. In this way, people's purchases no longer flock into those well-known top sellers and become more diverse and decentralized. We will further delve into the reputation system in IM based social commerce later in detail.

\begin{adjustwidth}{0em}{0em}
\textbf{Jump out of the box.}
\end{adjustwidth}

One interesting characteristic that makes IM based social commerce stand out is its capability of enabling people to ``jump out of the box", \emph{i.e.}, consider and thus further buy novel products that users are unlikely to try otherwise. In some cases, the buyers would not have even known the products without the recommendation from IM based social commerce:

\begin{adjustwidth}{2em}{2em}
\emph{"Last time I bought the snack of jujube with walnut kernels. This one was recommended by one of my acquaintances. Were it not her recommendation, I would not have known such combinations, not to mention buying it. ... She said she bought it, ... the jujubes were fresh and the walnuts were big ... I followed her ... bought a bag for each person in our family."} (P5)
\end{adjustwidth}

In some other cases, the buyers are stuck in finding products that can achieve certain functions, but did not know how to do so. Recommendations from IM based social commerce include experience from the socially-related ties. These experiences can be borrowed by users and users' horizons can be broadened, through which people access the products that they cannot think of before:

\begin{adjustwidth}{2em}{2em}
\emph{"Recently I bought a mosquito dispeller. Actually, before that I was looking for anti-mosquito liquid ... but was afraid that the chemical things are not good for our bodies. This (dispeller) works through physical mechanisms. I feel it less harmful to the body ... I had been trying to get a similar thing, but I didn't know a thing like this exists in the world and didn't think about it ... It now works very well. My house is now mosquito-free."} (P2)
\end{adjustwidth}

Through IM based social commerce, not only do users get to acknowledge new products, but they also give it a try to those products that they know but dare not try before. P4 expressed her experience of purchasing beef ribs in this way:

\begin{adjustwidth}{2em}{2em}
\emph{"One mom said she had tried this and it was very delicious. At that time I did not place an order, because I did not know how to cook it. But then I saw others follow her and show (the beef ribs they cooked) ... very attractive ... full of fragrance ... then I also had a try, found it particularly good, and placed one more order (laugh)."} (P4)
\end{adjustwidth}


Overall, exploiting the unique channel provided by IM, IM based social commerce makes more products reachable, and most importantly, worthy of trying and buying, to users. The same products can be accessed through a more direct and thus cheaper approach. Decentralized products whose brand names are not well-known but whose quality is good can be better accessed. Novel products that people have never known or unlikely to try are taken into considerations for purchasing. In these ways, IM based social commerce demonstrates an inclusive shopping experience.

\subsubsection{Time-wise}
Moreover, the IM feature of IM based social commerce has vastly strengthened users' reachability in the way that it 1) increases the timeliness of people's purchases and 2) induces longer-term and more willingly engagement.

\begin{adjustwidth}{0em}{0em}
\textbf{Timely.}
\end{adjustwidth}

On IM based social commerce, users are able to reach products in time. IM brings immediacy~\cite{nardi2000interaction}, providing near-synchronous communication and opportunities for timely information acquisition. What's more, people have already developed the habit of checking these IM applications which IM based social commerce embeds in frequently in their everyday lives. As such, sales information of IM based social commerce can more easily and more actively get into people's sights, and users can more efficiently acknowledge what products are concurrently for sales. This is particularly important when discounts for the products are available only for a limited amount of time:

\begin{adjustwidth}{2em}{2em}
\emph{"I have always wanted (to buy) perfume. You see, it is 99 yuan now ... But do you know how much did I pay for it, you know? 19.8 yuan. It was on discount only for a while. If you managed to seize the chance, you got it. If you didn't, it goes back to the original price." } (P1)
\end{adjustwidth}

What's more, timeliness is essential if the products belong to the category of scarce resources. For example, some participants mentioned they managed to buy masks through IM based social commerce when COVID-19 first broke out and masks were inaccessible through most channels:

\begin{adjustwidth}{2em}{2em}
\emph{"When the epidemic was at its worst, people couldn't buy masks. My mom happened to go to the bathroom at midnight and people in a sales group were recommending masks, urging everyone to rush for purchases. She made it."} (P6)
\end{adjustwidth}

As mentioned by P6, were it not the IM based social commerce sales group, people would find it hard to know which channels for these urgent resources are available at the time. The number of channels that one individual can monitor is relatively limited. However, combining the efforts of individuals and groups, people can more easily get to know information from more means. Through the power of word-of-mouth, the information is smoothly disseminated and people's needs are satisfactorily met.

\begin{adjustwidth}{0em}{0em}
\textbf{Ubiquitous.}
\end{adjustwidth}


IM based social commerce enables users to more casually get into the state of purchasing, turning the action of purchase more ubiquitous. With IM as the basis, it enjoys the welfare of IM's flexible nature~\cite{nardi2000interaction}. The messages would not take a long time for reading and the flexibility of IM lends people the privilege of deciding when to read the messages at their will: compared with face to face scenarios, IM requires neither temporal nor spatial co-presence. This enables people to digest promotion information at their own pace. What's more, compared with other platforms, an atmosphere of intimacy can be lent by the IM scenario~\cite{nardi2000interaction}. Therefore, promotions are less likely to be treated as a formal action because the promotions are just like \emph{"your friends are telling you something is good"} (P2). People's informal attitudes towards IM conversations are transferred to their perceptions of these sales at least to some extent, making the action of buying become more ubiquitous. Rather than scheduling in advance, users more willingly transfer into the status of purchasing:

\begin{adjustwidth}{2em}{2em}
\emph{"Generally I prefer getting into these groups in fragmented pieces of time. For example, when you are waiting for the elevator, or taking a break, you can take a look at it when you have time, or glance through it to see if there is anything good to buy recently."} (P6)
\end{adjustwidth}

In IM based social commerce, most products are recommended to exact items. People only need to judge if the items suit their taste and decide whether they intend to buy the recommended ones. With the lightweight nature of IM~\cite{nardi2000interaction} exploited, the purchase procedure is vastly simplified and therefore \emph{"can be easily finished in spare minutes"} (P8). Therefore, every time people do not need to devote much time to a single purchase decision, reducing the tedious procedure of comparing and judging and thus it leads people to have a stronger intention of retaining. In this way, the action of buying is turned from "bursty events", \emph{i.e.}, intentionally taking time to purchase, to "informal events", making quick purchases at people's will whenever free under a comfortable circumstance established by social relationships behind. The result of the accumulation of these spare minutes can be impactful:

\begin{adjustwidth}{2em}{2em}
\emph{"I look at it in my spare time. Our consumption capacity in leisure time is quite strong. For example, when looking at my cell phone before bed, I always want to buy something."} (P8)
\end{adjustwidth}

\subsubsection{Potential Downsides}
The integration of IM into commerce sometimes can lead to negative shopping experiences IM based social commerce. The informal and casual tenor~\cite{nardi2000interaction} of IM can sometimes be problematic when the mere communications and chats are turned to monetary issues. As expressed by P2:

\begin{adjustwidth}{2em}{2em}
\emph{"I buy more ... Probably the chances of regret are greater and maybe I buy more useless things ... more dispensable things ... But I never admit it (laugh)."} (P2)
\end{adjustwidth}

Through IM based social commerce, people have higher reachability to cheap, decentralized and "out of the box" products. However, when users' reachability to new products is enhanced, there is the possibility that these new products do not suit one appetite because "what you like may not be what I like" (P8). P1 referred to her experience of purchasing wild black wolfberries:

\begin{adjustwidth}{2em}{2em}
\emph{"I had only tried the red ones, but never the black one ... And they were recommending and buying, saying that wild black wolfberries are very tonic ... I took a risk (and bought it) ... But I have only drunk it once ... I am afraid that I might be poisoned."} (P1)
\end{adjustwidth}

However, most of the time, the purchase experiences on IM based social commerce are very positive. Through our study, we find that this phenomenon is closely related to the mechanism of human recommendation on IM based social commerce, which is to be discussed in the following.
\subsection{How People Make Purchase Decisions on IM based Social Commerce}
In answer to RQ~\ref{resques:3}, we find that the major feature that distinguishes IM based social commerce from prior social commerce is that IM based social commerce utilizes human recommendation in a trustworthy way. With IM as the basis, friends and real-life strong ties rather than algorithms or big brands/celebrities decide which products are recommended to users and promote the products. This integration of human in the process of recommendation brings plentiful benefits to IM based social commerce, which drastically shapes the decision-making process of making a purchase. Here, we first present findings on novel decision-making processes, then dive into mechanisms behind.

\subsubsection {Novel Decision Making Processes}
As emphasized by participants, the incorporation of IM has shaped the decision-making process of purchases to be more convenient and cost-reducing in terms of time and effort on IM based social commerce. This can work both in the scenario of passively receiving recommendations and actively searching for recommendations.

\begin{adjustwidth}{0em}{0em}
\textbf{Passively receiving recommendation.}
\end{adjustwidth}

Most recommendations on IM based social commerce are conducted in a passive manner (P2, P3, P10 \& P12). This is partly because on IM, others' messages can be reached without the constraints of temporal co-presence of the senders and receivers. As such, recommenders just need to constantly share links in IM (WeChat) groups, providing basic information and descriptions of the recommended items. When users of IM based social commerce want to see these links, they just enter the chatting interface and decide whether they would likely to purchase the items that are introduced. In this way, users of IM based social commerce feel that their purchase habits are remarkably altered. Compared to traditional commerce where in most cases users first \emph{"want to buy something and then search for the specific item"} (P10), IM based social commerce has changed this decision procedure. In IM based social commerce, people \emph{"first see what is being recommended, and then decide whether they need the item and whether it is worth buying"} (P10). 

Recommendations of this form make IM based social commerce recommendations ubiquitous, making the process of purchases more flexible and casual: 

\begin{adjustwidth}{2em}{2em}
\emph{"Let's say maybe I am free in the evening and I just look into these groups, go through them and see what they say. Sometimes I did not intend to buy, but their words may remind me that my home is also short of this, and then I buy it."} (P3)
\end{adjustwidth}

This ubiquity is regarded as bringing much convenience for purchases:

\begin{adjustwidth}{2em}{2em}
\emph{"You just need to browse the item she recommends. It enables me to save a lot of time if I buy through her channel."} (P1)
\end{adjustwidth}

\begin{adjustwidth}{0em}{0em}
\textbf{Actively searching for recommendation.}
\end{adjustwidth}

Some users also actively search for recommendations in IM based social commerce. This can be accomplished by the simple action of sending a message for consultation. Although as mentioned by participants that this only takes up a small share of their purchases on IM based social commerce, this can truly save people's time and effort in certain scenarios. For example, if users are not familiar with certain areas, they have the tendency of referring to the experienced for recommendations:

\begin{adjustwidth}{2em}{2em}
\emph{"If I want to please a little girl and give her a lipstick, but I don't know about it, I will ask [name], `what color should I buy'. It's when I want something but I don't know anything about the field, I ask an expert."} (P8)
\end{adjustwidth}

Similar circumstances are also noted by P4, P9, P10 and P11. For these people, actively searching for recommendations from friends who are trustworthy and better in certain areas can make up for their shortcomings in the corresponding aspects. This can not only save people's time for comparison, but also reduce the possibility of awful products. As a result, people's choice overload~\cite{moser2017no} is massively reduced and their purchase experiences are simplified.

Sometimes when experts are not available, users of IM based social commerce may turn to recommenders whose past recommendations are successful as substitutes.
For example, P1 described how she actively searched for suitcase through IM based social commerce recently:

\begin{adjustwidth}{2em}{2em}
\emph{"I was busy but my need was urgent ... I asked the leader of a sales group if there are 28-inch suitcases for recommendation ...  I had bought from him before, I came to him like a habit, totally unconsciously ... It is at least better than searching by yourself. How tiring it would be! ... He picks it for you. You save energy and time."} (P1)
\end{adjustwidth}

In these cases, users' needs are rather specific. Although these searches can be accomplished by themselves on traditional platforms, recommendations from IM based social commerce stand out in the way that it greatly diminishes the effort and time for product comparisons. IM based social commerce recommenders provide a relatively short list of products for consideration. With trust rooted in social relationships, people tend to believe in and accept their recommendations. As a result, the exhausting process of searching and comparing has been lessened to a simple inquiry through IM plus curtailed efforts on limited comparisons. This vastly reduces people's information overload, making the purchase decision process more convenient and improving users' purchase experiences.

\subsubsection{Drivers of Decision Making}

With IM delineating people's intimate relationships~\cite{nardi2000interaction} or even just "close friends, friends and sometimes family"~\cite{church2013s}, the ties on IM have the tendency of reflecting people's real-world social relationships. Highly integrated with these ties, IM based social commerce enjoys the welfare of social exchanges and thus interpersonal effects, which are caused by what Chinese people frame as 'guanxi'~\cite{yang2011virtual}. Influences aroused from close social relationships play a key role in making purchase decisions on IM based social commerce. Here, we analyze interpersonal effects from three angles: trust, homophily, and conformity.


\begin{adjustwidth}{0em}{0em}
\textbf{Peer to peer trust.}
\end{adjustwidth}

IM based social commerce significantly benefits from the mechanism of peer to peer trust. In the purchase decisions of IM based social commerce, while the items themselves demonstrate how much people need the items, social relationships with the recommenders determine to what extent it is worth having a try to follow the recommendation. If the recommender is someone that users are familiar with in the real world, they are more inclined to buy the items:

\begin{adjustwidth}{2em}{2em}
\emph{"In this [WeChat group name] group, every one of us is familiar with each other and knows each other well ... If one pushes something to me, I just buy without hesitation."} (P2)
\end{adjustwidth}

This is especially prominent when the number of people in the groups is rather small and limited to close ties in the real world. In cases like this, the closeness of people's social relationships induces trust within them. When these social relationships are integrated with recommendations in IM based social commerce, the innate trust lying within these social ties is transferred to the recommendations, making the recommended items seem worthy of a trial.

In other cases where the size of the group increases, scenarios where the recommender is not among the socially-close ones occur. However, if users acknowledge that someone familiar has bought it and enjoys it, they definitely are willing to follow the action of purchase:

\begin{adjustwidth}{2em}{2em}
\emph{"If you know someone close to you, or at least someone you know, had bought it, if they spoke highly of the product, I think I can buy it with no worries. I believe in them. There is no need for them to purposely deceive me."} (P6)
\end{adjustwidth}

From the perspective of consumers, these people's words act as past users' reviews. However, what makes these reviews stand out from the ordinary reviews on traditional e-commerce platforms is that the review sources are recognized as accountable:

\begin{adjustwidth}{2em}{2em}
\emph{"Basically, because we all know each other better, we know about the quality of their judgment or something. Therefore, as soon as she says something is good, we will follow her. Because we usually talk a lot, we know each other well."} (P4)
\end{adjustwidth}

In traditional commerce, users are hesitated to try items unless there are tens of thousands of reviews and the few drawbacks narrated in negative feedbacks are acceptable because they are cautious of the case of "click farming" (P10). However, these reviews on IM based social commerce are provided by "real persons" (P10), especially the ones that users know for sure. When the reviewer changes from an unfamiliar stranger to someone in users' social cycles, people's trust rooted in social ties makes every single review trustworthy, which greatly enhances the perceived value of every review. For example, one single review can lead to a purchase: \emph{"if one of my friends says it is good, I will definitely buy it"} (P1).
 
The integration of these peer-to-peer trust significantly optimizes the purchase decisions on IM based social commerce, saving people's time and effort:
 
\begin{adjustwidth}{2em}{2em}
\emph{"I believe in my friends' feedbacks more. I don't even want to think for a bit. I just buy it. I don't even read the comments. I want it, I like it, I just buy it and my time is saved. What a relief!"} (P2)
\end{adjustwidth}

\begin{adjustwidth}{2em}{2em}
\emph{"You will save a lot of time on searching and reading reviews, especially if it has been verified by others."} (P4)
\end{adjustwidth}

What's more, the existence of these known people makes the purchase process real and sincere. This leads people to be more willing to make the deal:

\begin{adjustwidth}{2em}{2em}
\emph{"You feel that these people are real, unlike traditional e-commerce where there are only some electronic codes, codes with positive praises: they are cold, with no temperature. But if the people around you tell you it is delicious, you will feel it is really delicious."} (P10)
\end{adjustwidth}

However, although not frequently mentioned, there would be circumstances where conflicting feedbacks are shared by different people. When this happens, some practitioners judge the reviews based on familiarity with and prior knowledge about the feedback providers:

\begin{adjustwidth}{2em}{2em}
\emph{"An important factor is how familiar I am with the two people ... If I am more familiar with someone, I would trust him more ... If one is an expert, I generally would believe her; while if the other is a novice online shopper, I may believe her to a lesser degree."} (P7)
\end{adjustwidth}

With IM social relationships that resemble real-world social ties, users have more background knowledge about the recommenders and thus the relative trustworthiness of their words. Therefore, decisions can be made according to the information provided by the relationships. Other practitioners may just emphasize more on the negative side, where they \emph{"lean on the poorly rated side"} (P10) or even \emph{"don't buy it as long as someone has a bad review"} (P2) because \emph{"she must have a reason to rate it bad"} (P10) and \emph{"there are so many choices and substitutes"} (P2). {\color{blue}These findings echo prior works \cite{sharma2013friends,sharma2013social} on social recommendations.}

\begin{adjustwidth}{0em}{0em}
\textbf{Trust of the community.}
\end{adjustwidth}

In some large WeChat groups where IM based social commerce works, it is not guaranteed that strong ties which induce strong peer-to-peer trust exist. However, despite the absence of these strong ties, the mechanism of trust can also work on the community level. Specifically, community mechanism and shared identity within the community can enhance people's trust of the community, which drives people to treat the items as buyable.

\emph{\textbf{a. Community mechanism.}} The mechanism of the community where recommendations of IM based social commerce occur contributes much to people's trust towards the community and thus people's willingness to buy items from the community.

Report mechanism is well developed in some sales groups for IM based social commerce. To be specific, users "@" the recommenders or at least notify them when they place an order, and report how the item is when the item arrives (one example of this is presented in Fig. \ref{fig:interface2}):

\begin{adjustwidth}{2em}{2em}
\emph{"Usually when I buy something, let's say she recommends (an item). I will tell her what I have bought in the group, saying (I have) `bought it'. If it is good after we purchase, we would report `it is good'."} (P2)
\end{adjustwidth}

Therefore, when this mechanism becomes routine and when other users can see the reports, they know 1) how popular the items are and 2) how good the items are, which may potentially determine one's judgment on to what extent the items are worth buying:

\begin{adjustwidth}{2em}{2em}
\emph{"They bought this and say it's good. They take a picture and send it out. I see (the picture), and find it really good ... Some may say this is no good. It under-weights. We all report this. And we generally believe that ... maybe we don't know everyone (who reports) ... I trust what they say."} (P3)
\end{adjustwidth}

Although a similar mechanism also exists in traditional commerce, these reviews are regarded as more credible. When asked why these reviews from strangers are deemed to be credible, some people refer to the mechanism of the construct of these groups where the transfer of trust plays a vital role. As mentioned by P3, in most cases although one does not know others in the groups directly, they may be one's friend's friend or one's friend's friend's friends. \emph{"Familiarity is transferred from layer to layer ... you feel assured"} (P3).

Norm mechanism is also identified as beneficial to trustworthiness. With well-established norms, the recommendation behaviors can be better regulated. This leads the experience in the community/group to be more enjoyable, and makes the recommendations seem more reliable. For example, some people regard the community's norm to deal with conflicts as praiseworthy:

\begin{adjustwidth}{2em}{2em}
\emph{"Sometimes disagreements arise. Some think an item is good, some think it's bad. On that occasion there are disagreements. Later, we say in our group, if there is a disagreement, we should not recommend it again."} (P1)
\end{adjustwidth}

With these norms developed, the remaining products recommended are ensured to be conceived as good by most existing purchasers. This leads the recommendations to be more robust, resulting in a higher degree of trust.

Some buyers in IM based social commerce also mention human feeling mechanism as a factor contributing to people's trust. Vastly entangled with social issues, the recommendations and the whole sales community would be regarded as dependable and thus buyable if the recommenders lend an atmosphere of human feeling to the community:

\begin{adjustwidth}{2em}{2em}
\emph{"(The recommender) sometimes buys herself. She tells you which product is good ... which item's discount is over and don't buy it ... For something new, she also asks the manufacturer how to use it ... She is very thoughtful. She is very enthusiastic and has a warm heart."} (P2)
\end{adjustwidth}

When a recommender thinks from the buyers' perspective, the recommendations act more like a sincere friend's caring. This leads people to regard the transaction as less commercial, resulting in people's enhanced trust and increased willingness to make purchases.

\emph{\textbf{b. Shared identity.}} When strong ties are not available, users of IM based social commerce regard the identities that lie underneath as the alternative resources for trust. Although users may not know someone in the real world, if he shares the same identity with them, trust is likely to yield. The identity can be a common status, or simply geographical similarity:

\begin{adjustwidth}{2em}{2em}
\emph{"Some parents of my son's classmates are selling things ... Because they (my son and their sons/daughters) are all classmates, you may have a higher degree of trust towards them. After all, everyone is together. They can't always lie to you."} (P3)
\end{adjustwidth}

\begin{adjustwidth}{2em}{2em}
\emph{"In fact, everyone doesn't know each other. But maybe because we belong to a small circle in common, for example, we live in the same neighborhood, or we are in the same area, we have a higher sense of identity."} (P8)
\end{adjustwidth}


As mentioned by buyers on IM based social commerce, with the assumption that 1) the weak ties where IM based social commerce builds upon should sustain, 2) people are identifiable through the shared identity, or 3) people should cherish the shared identity and should not stain the reputation of the shared identity, they tend to regard these people with shared identity with them as trust-worthy. With these shared identities as guarantees, people's fear of being deceived is mitigated. The results of the purchases are satisfying: \emph{"it's true that the quality is good after buying (in this way)"} (P3).

What's more, the more provable and more identifiable the shared identity is, the more likely that users of IM based social commerce enjoy a higher level of trust. For example, P1 regards the parent group as trust-worthy:

\begin{adjustwidth}{2em}{2em}
\emph{"We basically know the people in the group. ... Sometimes I am not very familiar with someone, but I am sure of his presence and he is very dependable. Because we use a real-name system in the group. When we get into the group, our identities have been verified. Although I can't remember so many people, the people in this group must be reliable."} (P1)
\end{adjustwidth}

As introduced by IM based social commerce  users, the introduction of identifiable real-name systems vastly enhances people's reliability. Sometimes they \emph{"cannot remember his name, but just know it is him"} (P1) and trust him.

However, circumstances like this mostly happen when the groups are not established specifically for transactions. When the groups are only for sales, it is relatively hard to introduce a real-name system. Nevertheless, if the entrance approval is only limited to existing members with identifiable information such as real names, the groups are still recognized as dependable and users think \emph{"there is no deception there"} (P2).

These forms of trust greatly shape people's decision-making process in purchases through IM based social commerce, bringing homophily and conformity to sight.

\begin{adjustwidth}{0em}{0em}
\textbf{Homophily.}
\end{adjustwidth}

Homophily stands out in the decision of choosing which channel of IM based social commerce to make the purchases. Users of IM based social commerce tend to rely more on those groups where a higher degree of homophily is shared for purchases:

\begin{adjustwidth}{2em}{2em}
\emph{"When the social circles of the group are different, the levels are different ... (I prefer) groups where people's levels are similar. Our income, background, and consumption level have something in common ...  sharing the same evaluation and a common expenditure on needs."} (P1)
\end{adjustwidth}

\begin{adjustwidth}{2em}{2em}
\emph{"I feel that what [name] recommends suits my taste ... You have to have something in common. If the things he recommends ... are going beyond rationality, I won't be with the group. ... Maybe this relates to what you focus on. A 20-year-old girl won't join our group. She won't be interested in what is pushing to me."} (P2)
\end{adjustwidth}

One main characteristic that makes IM based social commerce so unique and drives people to participate is the belief that human recommendation knows people better than machines. Therefore, how well the recommendations match people's tastes is crucial to the conclusion of transactions. Homophily leads to similarities, contributes to a higher probability of shared interests and thus a higher tendency to make the purchases.

What's more, as mentioned by participants, people's level of consumption determines whether they regard buying a product at a price is worth it, while their aesthetic tastes decide whether a product is worth buying. Therefore, purchases are more likely to be led to when a high level of homophily is shared in these two aspects, for example, mom groups at a high school (P1, P2 \& P4) and student groups at certain universities (P12). With relatively similar purchase capabilities and tendencies as backings, these communities have a higher likelihood to sustain.

\begin{adjustwidth}{0em}{0em}
\textbf{Conformity.}
\end{adjustwidth}

Conformity stands out in the decision of whether to purchase an item. As noted by participants, a great many items users buy on IM based social commerce platforms may be "dispensable" (P1) or at least "substitutable" (P6). However, if someone spots that many people are purchasing something, he or she has the innate impulse to follow others. This frequently happens within IM based social commerce, especially when groups for IM based social commerce also serve the function of aggregating the homophilous: the existence of a sales group enables users to more easily acknowledge what others akin to them are buying. When much similarity is shared and others are regarded as trustworthy, a higher tendency of conformity is likely to be induced:

\begin{adjustwidth}{2em}{2em}
\emph{"If an item appears in the group and many people in the group sequentially buy it, then you will also buy it and think, this product should be so good, everyone is rushing towards it ... it must have its advantages, then I will try it too. The most delicious Kimchi my mom bought was brought back in this way."} (P8)
\end{adjustwidth}

If everyone who has purchased the items speaks highly of them, IM based social commerce buyers can enjoy a higher level of willingness to conform, where the power of Word of Mouth is well-utilized:

\begin{adjustwidth}{2em}{2em}
\emph{"Recently I bought the Mutton Vermicelli. A person from Guizhou says it is the taste of his hometown and shared (a photo) ... Several people followed him, and report with pictures, saying it is very delicious. This began to spread within the group and made people 'panicky'. Anyone who eats mutton wants to buy it."} (P2)
\end{adjustwidth}

In these cases, the WeChat groups which IM based social commerce embeds in serves as a semi-open reliably community. Several dependable people simultaneously and spontaneously buy or praise something, virality can be fermented: more and more people around in the group take a try, and people's inclination of following the purchased is increasing bit by bit.

What's more, conformity can shape people's trust in turn. If the items that are bought through following others' purchases are satisfying, people are more likely to believe their judgments and thus their trust of the group and community can be enhanced:

\begin{adjustwidth}{2em}{2em}
\emph{"If I follow other's purchases and it turns out well, I will definitely trust them more the next time."} (P4)
\end{adjustwidth}

\subsection{How IM Based Social Commerce Influence Social Relationships}


While social relationships profoundly shape IM based social commerce, it is intriguing to discover whether this holds vice versa, \emph{i.e.}, whether the integration of IM based social commerce into people's social lives can also affect their social ties (RQ~\ref{resques:4}). Based on what has been mentioned by IM based social commerce users, we discover that 1) IM based social commerce can cultivate the formation of new ties; 2) IM based social commerce can strengthen weak ties; and 3) for strong ties, IM based social commerce brings no changes, where overuse can harm existing ties, but the damage is limited. 

\subsubsection{Forming new ties and strengthening weak ties.}
Good use of IM based social commerce can benefit social relationships, establishing new relationships and enhancing existing weak relationships. For example, P1 mentioned she was not acquainted with an IM based social commerce recommender in the beginning, but they became friends afterward due to their frequent interactions on IM based social commerce:

\begin{adjustwidth}{2em}{2em}
\emph{"Her daughter went to the same high school as my son ... But I did not even know her daughter in the beginning ... Without her, I would have bought at a higher price ... I now add her as my friend."} (P1)
\end{adjustwidth}

As a channel for informal lightweight communication, the informal and friendly tenor of IM bring intimacy~\cite{nardi2000interaction}. This sense of intimacy can sometimes be transferred towards the socio-economic interactions of IM based social commerce when certain characteristics are felt, for example, warmth and earnestness (P1). When people are interacting not only socially but also financially, a higher degree of trust and closeness is likely to be established if the purchase experiences are satisfying. As such, new ties tend to be thus formed and weak ties tend to be thus strengthened. 

Recommendations and reviews on IM based social commerce can also serve as themes to discuss with others. With more communications concerning purchases, people have more common topics to talk about, and their interaction frequency increases. This can consequently improve social closeness, especially when the ties are not so strong in the beginning:

\begin{adjustwidth}{2em}{2em}
\emph{"Mostly the social relationships are enhanced ... If we both buy bread (from the same source), we would talk about the bread ... If I have a good approach to shopping, I will also recommend it to them. This will increase the chance to communicate with each other, and our relationship will be closer."} (P10)
\end{adjustwidth}

What's more, when users generate more good recommendations or reviews, they are regarded as more trustworthy and people's relationships with them are perceived to be enhanced: \emph{"if anyone can recommend good things, he will have some kind of prestige. I will believe in him"} (P2).

\subsubsection{Maintaining strong ties.}

In most cases, people perceive IM based social commerce brings no changes to people's existing strong social ties. Some people clearly differentiate the groups for purchases and for social interactions. When regarding some groups as specifically for buying, people tend to have more tolerance for the frequent messages in the group. Sometimes to avoid disturbance, people just set the groups to disturb-free or block others' WeChat moments:

\begin{adjustwidth}{2em}{2em}
\emph{"I turn the groups to disturb free. ... The reason why we get into the group is to buy from it ... I haven't spotted cases where relationships are affected."} (P3)
\end{adjustwidth}

In other cases, social relationships and transaction relationships are intertwined. However, people distinguish clearly between the annoyance of people's behavior and favor of social relationships:

\begin{adjustwidth}{2em}{2em}
\emph{"If we are very close and I post (ads/recommendations) every day, my WeChat moment has nothing about my life, you may also block me. ... But this has nothing to do with our associations with the person."} (P11)
\end{adjustwidth}

Some people attribute this to empathy:

\begin{adjustwidth}{2em}{2em}
\emph{"You just need to understand them ... It's also a means of survival, a normal behavior. ... In fact, a lot of my friends do business now. More or less, we help a bit. Isn't everything you do is a matter of survival?"} (P8)
\end{adjustwidth}

As mentioned by IM based social commerce buyers, if users are close enough to the friends who now become recommenders of IM based social commerce, they have some sort of understanding and consideration for their actions of sales promotion for living. Therefore, people seem to tolerate the information, and sometimes even buy some of the products to support them.

Although not very frequently mentioned, when the recommendations occur much too often and too unwillingly, existing strong ties can be hurt: 

\begin{adjustwidth}{2em}{2em}
\emph{"One of my friends recommends make-ups ... We don't like cosmetics, but she is always promoting it to you and getting you to buy ... That is kind of annoying ... When we see other friends we would say, why is she always doing that."} (P3)
\end{adjustwidth}

However, the damages are mostly rather limited both in terms of extent and duration. If the ties are backed by long-term solid social interactions, the consequence of the damages is relatively slight. When there are no longer disturbing actions, strong ties can be restored as they were. Just as mentioned by P3 about the same person:

\begin{adjustwidth}{2em}{2em}
\emph{"It does not hurt much. We went to the same kindergarten, elementary school, and junior high school. We just ignored her ... Gradually she seemed aware (of the overuse), and stops talking about it ... Our relationships become fine as before."} (P3)
\end{adjustwidth}
\section{Discussion}

Based on findings from our study, we proceed to discuss the research and design implications, and show the limitations and future work.

\subsection{Making the Weak Strong}

First of all, we demonstrate that IM based social commerce significantly deviates from traditional commerce in its decentralized nature, where the major drivers of the platform are ordinary people~\cite{cao2019your} and long-tailed products. 

\textbf{Actors on IM.} While traditional commerce and marketing largely depends on key opinion leaders such as celebrities and big brand to broadcast, IM based social commerce rely on ordinary people to do the marketing, as we show in Section 4.1. While individuals on IM based social commerce can only exert local influence and is much weaker compared to key opinion leaders, the aggregated power of such individuals is tremendous -- here people recommend products to their friends and families, and the recommendation information quickly diffuses over social networks. Moreover, as the network is built on existing social ties, the recommendation appears much more trustworthy (which in turn leads to purchase) compared to the advertisement on traditional commerce/traditional social commerce, which has been mentioned and highlighted by many participants {\color{blue}especially in terms of Section 4.3.2}. Essentially, due to this mechanism, IM based social commerce turns the traditional `weak' individuals into great economic value. 

\textbf{Products on IM.} On the product level, IM based social commerce also demonstrates that it is `making the weak strong'. In traditional commerce, people typically buy popular products that are well-advertised~\cite{yadav2013social} -- typically products from well-known companies that are of power and enough resources, and it is hard for less well-known products to get noticed by people. Meanwhile, we find that social commerce helps long-tailed products prosper. {\color{blue}As revealed in Section 4.2.1,} they are more likely to purchase less well-known brands compared to their experience in traditional commerce, often per friend's recommendation. 

In summary, we argue that IM based social commerce demonstrates the unique characteristic of `making the weak strong', and it revolutionizes the way resources and power get allocated, making things more decentralized and turning user experiences to be more inclusive.



\subsection{Re-Blending Social and Economic Lives}
Moreover, IM based social commerce significantly changes users' shopping and social experience. Traditionally, people's social and economic lives are rather intertwined \cite{granovetter1985economic}. However, with a growing trend of labor division, the functions of different objects become focused and specialized. The advent of globalism and e-commerce has facilitated this differentiation and disentanglement of social ties and economic ties, separating social lives and economic lives to be independent~\cite{granovetter1985economic}. Socially, platforms such as social media serve the role of relationship maintenance. While for consumption, people head to specialized e-commerce sites, getting recommendations largely from recommender systems and reading reviews posted by strangers. IM based social commerce, however, is one way to re-blend social and economic lives together back in the new era. Through informal and lightweight communication, purchases and recommendations take place on the same medium, IM, where people socialize with their friends. With the expressive and flexible context of IM~\cite{nardi2000interaction}, {\color{blue}as we reflect in Section 4.2,} shopping through this means becomes informalized and more ubiquitous. People get trusted recommendations from friends, easily transfer into the state of purchase and engage in shopping without intentionally scheduling a shopping time. Meanwhile, shopping can influence users' social lives as well. As discussed in Section 4.4, people get to know new friends or maintain closer relationships with existing friends as they interact on IM based social commerce -- of course, relationships may be negatively influenced by economic behavior under certain situations, but the consequences are often limited. Such a blend of social and economic lives lets users engage longer time on the platform, and essentially changes the way people purchase and socialize through technology. 





\subsection{Design Implications}

Our work provides novel design implications from several aspects. Through the interview study, we demonstrated the benefit of incorporating human issues into the decision loop. Compared to the traditional e-commerce platforms where people individually enter the sites and search for products to buy, on IM based social commerce humans, who are real-world friends, send the recommendations. Consequently, as we show in Section 4.3, plentiful benefits can be brought by the social exchanges and interpersonal influences with human recommendations from social connections. The power of trust may be magnified in some communities. 
The human feeling mechanism would not have been possible if it were not the inclusion of real humans. Hardly would a sense of shared identity be felt without establishing explicit inclusive IM communities. Therefore, we advocate that future design of the kind should better consider how to include these intangible but impactful human issues into platform design so as to engage users and make the process more trustworthy and inclusive. {\color{blue}For example, human-computer cooperative recommendations and decisions would be a feasible trial.}

Our study offers concrete design implications for IM based social commerce as well. While friends' recommendations play a key role in IM based social commerce, the current implementation is rather straightforward: friends directly share links in IM group or post through direct message. Recall that Section 4.3 illustrates how people's decisions can be shaped by both individual friend's and community's purchases (conformity) and reviews (peer-to-peer trust). {\color{blue}Inspired by Terveen et al.~\cite{terveen1997phoaks}, we believe for future design, social commerce can potentially implement an interface where users can check features such as (1) their friends' buying history, (2) their friends' ratings and reviews on items, (3) the buying history of certain communities, and (4) a community's overall ratings and reviews on items so as to decide what they want to buy. Some initial steps have been taken to incorporate the circumstance of (1) in some social commerce platforms, but we regard (2)-(4) also as doable, which we believe would be a fine complement to the current model. In essence, such model would be a combination of algorithm-driven e-commerce (traditional e-commerce) and social commerce that potentially leverage strengths from both sides: algorithm automatically captured the community level (in contrary to traditional e-commerce that aggregates reviews across a whole platform's worth of users) information that is most likely to capture the user's most likely preference in the trusted social group.}  Of course there can be potential privacy issues, thus the design should offer users to enable/disable the checking of friend's behavior and give people the right to decide whom to share with, mimicking that of determining which IM groups to distribute to. 

Our findings also demonstrate that there can be potential downsides of social commerce, where over-exploiting social relationships for economic purpose can pose negative impacts on social relationships, which the business model relies on. Thus, the platform and recommenders on IM based social commerce should be especially aware of the extent of marketing they carry out online. {\color{blue}Future research should address the extent of `healthy' social recommendation and social commerce site should encourage users to follow the corresponding best practices -- future interface of social commerce could potentially add a `warning' feature to recommenders when they over market and may potentially hurt their social relationships.}

Our research further contributes to building decentralized and trusted platforms in general. {\color{blue}In line with prior works~\cite{cao2019your,chen2020intermediary}, we demonstrate that contrary to prior social commerce where Key Opinion Leaders play a vital role (e.g., Facebook commerce, Twitter commerce), the major actors on IM based social commerce platforms are ordinary people and their strong ties, and we show how influence can be exerted by these `decentralized nodes', which results in trusted recommendations/buying relationships.} Notably, all of the relationships and mechanisms are extensions of the pre-existing social network on IM, thus the system is much more cost-efficient to build and much easier to attract and engage users compared to prior commerce since there is no need to build the platform from scratch. Moreover, IM makes the weak strong and helps diffuse traditionally under-representative innovations, making the information diffusion process more inclusive. Based on the success of IM based social commerce, we envision the possibility of aggregating social capital resources through existing IM platforms to achieve other collective goals in a similar fashion, which would be especially useful in cases where trust plays a key role, for example, leveraging IM to promote crowdsourcing and crowdfunding~\cite{althoff2015donor}.



\subsection{Limitations and Future Work}
Our study also suffers from several limitations. First of all, our data may suffer from selection bias. Users with positive experiences of IM based social commerce may be more likely to participate in our research compared to those that do not. Moreover, user experiences may be influenced by social demographics (e.g., gender, age, income) and social power. While we tried to recruit users from diverse backgrounds, given the great coverage of IM based social commerce, we cannot ensure all cases will be covered by our study. Future work should carry out larger-scale analysis, or verify the findings through empirical data collected from the social commerce side. 

\section{Conclusion}
With a semi-structured interview study on the buyers of instant messaging (IM) based social commerce, we investigate how and why people engage in these platforms. Through qualitative analysis, we discover that social relationships play a vital role in people's involvement {\color{blue}in} IM based social commerce. Users' enhanced reachability both product-wise and time-wise distinguishes IM based social commerce. The utilization of human recommendation brings interpersonal influences including peer to peer trust, community trust, homophily, and conformity {\color{blue}to} stand out in IM based social commerce and reduces the cost needed for the decision making process for purchases. The integration of IM based social commerce into social lives can bring changes to social ties. These discoveries can indicate future research and design for social-computing based marketing platforms and applications of computer-supported cooperative work.

\begin{acks}
The authors thank Fengli Xu for helpful suggestions on polishing the work. This work was supported in part by The National Key Research and Development Program of China under grant 2020AAA0106000, the National Natural Science Foundation of China under U1936217,  61971267, 61972223, 61941117, 61861136003, Beijing Natural Science Foundation under L182038, Beijing National Research Center for Information Science and Technology under 20031887521, and research fund of Tsinghua University - Tencent Joint Laboratory for Internet Innovation Technology.
\end{acks}

\bibliographystyle{ACM-Reference-Format}
\bibliography{refer}

\end{document}